\documentclass[aps,amsmath,amssymb, reprint, prl, preprintnumbers]{revtex4-1}
\pdfoutput=1 

\usepackage{graphicx}
\usepackage{epstopdf}
\usepackage{amsmath, amssymb}

\def\IZ{\mathbb{Z}}

\def\CN{{\cal N}}

\def\half{\frac{1}{2}}

\def\Tr{{\rm Tr}}

\begin{document}

\title{Supersymmetric Spectral Form Factor and Euclidean Black Holes}

\preprint{KIAS-P22047}

\author{Sunjin Choi}
\affiliation{School of Physics, Korea Institute for Advanced Study,
85 Hoegiro, Dongdaemun-gu, Seoul 02455, Republic of Korea.}
\email{sunjinchoi@kias.re.kr}

\author{Seok Kim}
\affiliation{Department of Physics and Astronomy \& Center for Theoretical Physics,
Seoul National University, 1 Gwanak-ro, Seoul 08826, Republic of Korea.}
\email{seokkimseok@gmail.com}

\author{Jaewon Song}
\affiliation{Department of Physics, Korea Advanced Institute of Science and Technology,
291 Daehak-ro, Yuseong-gu, Daejeon 34141, Republic of Korea.}
\email{jaewon.song@kaist.ac.kr}

\begin{abstract}

The late-time behavior of spectral form factor (SFF) encodes the inherent discreteness of a quantum system, which should be generically non-vanishing. We study an index analog of the microcanonical spectrum form factor in four-dimensional $\mathcal{N}=4$ super Yang-Mills theory. 
In the large $N$ limit and at large enough energy, the most dominant saddle corresponds to the black hole in the AdS bulk. This gives rise to the slope that decreases exponentially for a small imaginary chemical potential, which is a natural analog of an early time. We find that the `late-time' behavior is governed by the multi-cut saddles that arise in the index matrix model, which are non-perturbatively sub-dominant at early times. These saddles become dominant at late times, preventing the SFF from decaying. These multi-cut saddles correspond to the orbifolded Euclidean black holes in the AdS bulk, therefore giving the geometrical interpretation of the `ramp.' Our analysis is done in the standard AdS/CFT setting without ensemble average or wormholes. 
\end{abstract}


\maketitle

\section{Introduction}
One form of the information paradox in quantum gravity involves the long time behavior of correlators \cite{Maldacena:2001kr} or more simply the spectral form factor (SFF) defined as \cite{Papadodimas:2015xma}
\begin{align} \label{eq:SFFcanonical}
Z(\beta+it)Z(\beta - it) = \sum_{n, m} e^{-\beta(E_m+E_n) + i t(E_m-E_n)} \ . 
\end{align}
It probes a finer detail of the system for large $t$, and it should not vanish at late times (large $t$) for a typical quantum mechanical system with a discrete spectrum. However, in the smooth geometric description of gravity, one finds that the spectral form factor (SFF) decays rapidly at early times ($t$ small) and the main contribution comes from the black hole. The challenge is to find a geometrical description that accounts for the correct late-time behavior of the SFF. 

Recently, there has been significant progress in understanding such geometric description that accounts for microscopic information \cite{Cotler:2016fpe, Saad:2018bqo, Saad:2019lba, Penington:2019npb, Almheiri:2019psf, Almheiri:2019hni, Almheiri:2019qdq}.
Especially low-dimensional systems such as JT gravity and the SYK model have been extensively studied. There the apparent puzzle in the bulk description of the SFF is accounted for by the spacetime wormholes, connecting two disconnected spatial regions, whose contribution stops the SFF from decaying. However, this solution gives rise to another puzzle since the setup involves ensemble averaging, even though most of the top-down construction of AdS/CFT correspondences does not require any ensemble averaging. Therefore it is certainly desirable to ask if there is a way to account for the `information loss' in a conventional AdS/CFT setting such as $\CN=4$ Super Yang-Mills theory (SYM) that does not require ensemble averaging or wormholes. 

To this end, we study a variation of the spectral form factor that can be computed reliably in the supersymmetric field theory without ensemble averaging. Instead of the thermal partition function, we consider an index version of the partition function, with $(-1)^F$ inserted. It can be computed even at the strong-coupling due to the supersymmetric localization or the deformation invariance. It is schematically given as \cite{Romelsberger:2005eg, Kinney:2005ej}
\begin{align} \label{eq:grand}
    Z(\mu) = \Tr_{\mathcal{H}} (-1)^F e^{-\beta \delta} e^{-\mu j} = \sum_j \Omega_j e^{- \mu j} \ , 
\end{align}
where $\delta = \{ Q, Q^\dagger \}$ is an effective Hamiltonian on $S^{d-1}$ ($d$ being the spacetime dimension) given by an anti-commutator of supercharges, $j$ is a charge that commutes with $\delta$ and $\mu$ is its chemical potential. The trace is over the states on $S^{d-1}$ and the index gets contributions only from the states with $\delta = 0$ due to the bose-fermi cancellation. Generally, there can be many more chemical potentials, but let us suppress them for the moment. 

We will regard the chemical potential $\mu$ and $j$ as an analog of the inverse temperature $\beta$ and energy $E$, and study the behavior of $Z(\mu + i \tau)$ for large `time' $\tau$. 
One may worry that the $(-1)^F$ insertion can result in bose-fermi cancellation, leading to severe under-counting of states. 
However, it has been shown recently that the index correctly captures the entropy of BPS black holes in AdS \cite{Cabo-Bizet:2018ehj, Choi:2018hmj, Benini:2018ywd} and exhibits Hawking-Page type phase transition as we vary the chemical potential \cite{Choi:2018vbz, Copetti:2020dil}. Hence, the supersymmetric index is an appropriate probe for the fine-grained nature of gravity, and it is natural to regard the chemical potential as an analog of the inverse temperature. 

It turns out the direct analog of \eqref{eq:SFFcanonical} does not decay fast enough for `early times.' Instead, we consider the microcanonical version of the SFF, defined as
\begin{align}
    Y_{j, \Delta} (\tau) = \sum_{\tilde{j}}   e^{-\frac{(\tilde{j} - j)^2}{2\Delta^2}} \Omega_{\tilde{j}} e^{-i \tilde{j} \tau}\ , 
\end{align}
where we have a Gaussian window around $\tilde{j}=j$ with width $\Delta$. 
The (non-index version of) microcanonical SFF for the free large $N$ Yang-Mills theory was studied in \cite{Chen:2022hbi}, where the role of various saddles has been shown to be important. 
Here, we study the large $N$ behavior of $Y_{j, \Delta}(\tau)$ with $j \sim O(N^2)$ and $\Delta \sim O(N)$ for the $\CN=4$ SYM theory. In this regime, contribution to the SFF at early time comes from the BPS black hole with charge $j$ \cite{Gutowski:2004ez}, which decays exponentially in $\tau$. 

This causes a version of `information paradox' since we know that the SFF should not go to zero for large $\tau$ \footnote{In fact, for the case of $\CN=4$ SYM theory, it is periodic since the relevant charges of the states counted by the index are quantized.}. For us, this is saved via what we call as $K$-cut saddles \cite{Hong:2018viz, Choi:2021rxi}. 
These saddles correspond to the $\IZ_K$-orbifolded Euclidean black holes in the bulk AdS$_5 \times S^5$ \cite{Aharony:2021zkr}. As we will see, the $\IZ_K$-orbifolded black hole contribution becomes dominant at large $\tau$ and SFF increases along a `ramp,' resolving the `information paradox.'

\section{Large $N$ saddle points of the superconformal index of $\CN=4$ SYM}
In this paper, we focus on $\CN=4$ supersymmetric Yang-Mills theory with $U(N)$ gauge group in the large $N$ limit. The superconformal index of this theory is defined as \cite{Romelsberger:2005eg, Kinney:2005ej} 
\begin{align}
Z(\Delta_I,\omega_i) = \textrm{Tr} \left[ (-1)^F e^{-\sum_{I=1}^3 \Delta_I R_I -\sum_{i=1}^2 \omega_i J_i } \right]\ ,
\end{align}
where the trace is taken over the Hilbert space of radially quantized SCFT on $S^3 \times \mathbb{R}$. Here, $R_I$'s and $J_i$'s are the Cartans of the $SO(6) \cong SU(4)$ R-symmetry and the $SO(4)$ rotation symmetry respectively. In our convention, the chemical potentials satisfy the constraint
\begin{align}\label{con}
\Delta_1+\Delta_2+\Delta_3 - \omega_1 - \omega_2 =  2 \pi i \ .
\end{align}
The index counts the $\frac{1}{16}$-BPS states satisfying $E=R_1+R_2+R_3+J_1+J_2$, where $E$ is the energy. The index is invariant under any continuous marginal deformations. The index for the $U(N)$ $\CN=4$ SYM theory can be written as a matrix integral given as \cite{Kinney:2005ej}
\begin{align}\label{contour-index}
\begin{split}
& Z(\Delta_I,\omega_i) = \frac{1}{N!}  \prod_{a=1}^N \int_0^{2\pi} \frac{d\alpha_a}{2\pi} \prod_{a<b} \left(2\sin \frac{\alpha_{ab}}{2}\right)^2 \\
&\times \exp \! \left[ \sum_{n=1}^\infty \frac{1}{n} \! \left(1-\frac{\prod_{I=1}^3(1-e^{-n\Delta_I})}{\prod_{i=1}^2(1-e^{-n\omega_i})}\right) \!\! \sum_{a,b} e^{in\alpha_{ab}} \right],
\end{split}
\end{align}
where the contour integral variable $e^{i\alpha_a}$'s are the gauge holonomies of the $U(N)$ gauge group and $\alpha_{ab} \equiv \alpha_a-\alpha_b$.

The integral representation of the index \eqref{contour-index} in the large $N$ limit can be evaluated using a saddle point analysis. The large $N$ saddle points of the index integral we consider are labeled by three integers $K, r_1, r_2$ \cite{Choi:2021rxi}. 
Here, $K \ge 1$ represents the number of cuts or clusters of densely distributed eigenvalue $\alpha_a$'s in the large $N$ limit. 
The value of the partition function for the equal R-charges $R_1 = R_2 = R_3$ at the saddle $(K,r_1, r_2)$ is given by \cite{Choi:2021rxi}
\begin{align}\label{free}
\textrm{log} Z^{K, \pm}_{r_1,r_2} \!=\! \frac{4N^2}{27 \prod_{i=1}^2\left(\omega_i\!-\!\frac{2\pi ir_i}{K}\right)} \left( \omega\!-\!\frac{2\pi i(r\! \mp\! \half)}{K} \right)^3 ,
\end{align}
where $\omega \equiv \frac{\omega_1+\omega_2}{2},r \equiv \frac{r_1+r_2}{2}$ \footnote{The choice of $\pm$ is related to the way of unrefining $\Delta_I$'s to a single chemical potential under the constraint \eqref{con}.}.

Among the above solutions, not all of $(K,r_1,r_2)$ give proper large $N$ saddle points. One should appropriately choose $r_1,r_2 \in \mathbb{Z}$ for given $\omega_{1,2}$ and $K, s=\pm$ so that the following conditions are satisfied \footnote{These conditions are weaker than the conditions derived in \cite{Choi:2021rxi} in general. They coincide when $\omega_1=\omega_2$. We believe it was due to the technical limitation of obtaining the large $N$ saddle points of the index. Here, these weaker conditions come from the dual AdS$_5 \times S^5$ gravity side: stability against the condensation of D3-brane instantons in the Euclidean BPS black hole backgrounds \cite{Aharony:2021zkr}. Also, they are equivalent to the on-shell conditions for the Lorentzian BPS black holes \cite{Gutowski:2004ez, Gutowski:2004yv, Chong:2005da, Kunduri:2006ek} to have positive entropy at positive charges with no closed time-like curve.}:
\begin{equation}\label{r-con}
\begin{aligned}
    &\pm \textrm{Im} \left(\frac{K\omega_j-2\pi i (r_j \mp 1)}{K\omega_i -2\pi i r_i}\right) > 0 \quad (i \neq j)\ , \\
    &\mp \textrm{Im} \left( \frac{(K\omega-\pi i(2 r \mp 1))^2}{(K\omega_1 -2\pi i r_1)(K\omega_2 -2\pi i r_2)} \right) >0\ .
    \end{aligned}
\end{equation}
When the ratio $c=\frac{\omega_2}{\omega_1}$ is real, the two conditions in the first line are greatly simplified as
\begin{equation}
    -1<s\, (c\, r_1 - r_2)<c\ ,
\end{equation}
which is independent of $\omega$ and $K$. In particular, when $c=1$, arbitrary integer $r=r_1=r_2$ is allowed. The condition in the second line non-trivially depends on $\omega$ even for this simple case, so it should be considered separately.

Let us remark that the $K$-cut saddle spontaneously breaks the $U(1)^{(1)}$ 1-form center symmetry of the $U(N)$ gauge theory to $\mathbb{Z}_K^{(1)}$. When $\omega_1=\omega_2$, in the microcanonical ensemble, the entropy of the $K$-cut saddle is given by $\textrm{Re}(S_K) = \frac{\textrm{Re}(S_{K=1})}{K}$ regardless of $r$ and $s$, so that the entropy coming from the $K>1$ are subdominant \cite{ArabiArdehali:2019orz, Aharony:2021zkr, Choi:2021rxi}.

When $\omega_1 = \omega_2$, the large $N$ saddle points parameterized by $(K,r)$ are in one-to-one correspondence \cite{Choi:2021rxi} with the $(K,r)$ Bethe roots discussed in \cite{Hong:2018viz, Benini:2018ywd, ArabiArdehali:2019orz, Aharony:2021zkr}, which preserve the $\mathbb{Z}_K^{(1)}$ 1-form center symmetry. In the dual gravity side, the solutions with $K=1$ correspond to the multiple Euclidean solutions \cite{Cabo-Bizet:2018ehj, Cassani:2019mms, Aharony:2021zkr} mapped to the same Lorentzian BPS black hole solution in AdS$_5 \times S^5$ \cite{Gutowski:2004ez, Gutowski:2004yv, Chong:2005da, Kunduri:2006ek}. When $K>1$, they correspond to certain supersymmetry preserving $\mathbb{Z}_K$ orbifolds on the temporal circle and five angular circles of the above Euclidean BPS black hole solutions in AdS$_5 \times S^5$ \cite{Aharony:2021zkr}. Namely, this orbifold acts on the cigar geometry of the Euclidean black hole solutions breaking the $U(1)$ winding number conservation symmetry of strings to $\mathbb{Z}_K$.

\section{Spectral Form Factor of $\CN=4$ SYM}
The supersymmetric index version of the spectral form factor in the large $N$ limit can be computed using multiple saddle point approximations (one for the free energy, the other for the inverse Laplace transformation). To simplify our problem, we shall only consider the case with equal R-charges: $R_1=R_2=R_3 \equiv R$. In addition, let us fix the ratio of $\omega_1$ and $\omega_2$ as $c \equiv \omega_2 /\omega_1$ and perform the Legendre transformation with respect to $\omega \equiv \frac{\omega_1 + \omega_2}{2}$. This gives us degeneracy $\Omega_j$ of the states with charge $j \equiv 6\left(R+ \frac{1}{1+c}J_1 + \frac{c}{1+c}J_2 \right)$. Using this, we can compute the spectral form factor as
\begin{align}\label{SFF}
    \begin{split}
        Y_{j,\Delta} & (\tau) = \sum_{\tilde j} e^{-\frac{(\tilde j-j)^2}{2\Delta^2}} e^{-i \tilde j t} \Omega_{\tilde j} \\
        & \sim \int d\tilde j e^{-\frac{(\tilde j-j)^2}{2\Delta^2}} e^{-i \tilde j t} \int_C  \frac{d\omega}{6\pi i} e^{\frac{\omega-2\pi i}{3}\tilde j} Z(\omega) \\
        & = \int  \frac{d\omega}{6\pi i} Z(\omega) \int d\tilde je^{-\frac{(\tilde j-j)^2}{2\Delta^2}} e^{\frac{\omega-i\tau}{3} \tilde j} \\
        & \sim \int d\omega \frac{\Delta}{3\sqrt{2\pi} i} Z(\omega) e^{\frac{\omega-i\tau}{3}j + \frac{(\omega-i\tau)^2}{18}\Delta^2} \\
        & \sim \sum_\star \exp\left[\frac{\omega_\star}{3}j + \frac{\omega_\star^2}{18}\Delta^2 + \log Z(\omega_\star+i\tau)\right] ,
    \end{split}
\end{align}
where we defined $\tau=3t+2\pi$ when we go from the second to the third line and $C$ is an appropriate contour parallel to the imaginary axis. Here, we omitted $c=\frac{\omega_2}{\omega_1}$ for simplicity. The last sum is over the $(K, r_1, r_2, \pm)$ as we discussed.

We assume $\tau$ to be real-valued and of order $\mathcal{O} (N^0)$ and $j>0$ to be of order $\mathcal{O}(N^2)$, $\Delta>0$ to be of order  $\mathcal{O}(N^1)$.
Then, $\Omega_{\tilde j}$ is of order $\mathcal{O}(e^{N^2})$ in the region $\tilde j \in (j-\Delta,j+\Delta)$ as we have discussed. Here, $\omega_\star = \omega_\star (\frac{j}{N^2},\frac{\Delta}{N},\tau)$, which is of order  $\mathcal{O}(N^0)$, denotes the large $N$ saddle points of the $\omega$ integral that give $\log Z(\omega_\star +i\tau) = \mathcal{O}(N^2)$. The saddle point equation is given by a quintic equation of $\omega$ in general and we should only take the solutions satisfying \eqref{r-con} with $\textrm{Re}(\omega_\star)>0$.

Let us analyze the behavior of $Y_{j,\Delta}(\tau)$.
Namely, we consider 
\begin{align}
\hspace*{-0.15cm} Y_{j,\Delta} \!\sim\! \sum_{\substack{K \in \mathbb{N} \\ s=\pm \\ r_{1,2}}} \! \exp\left[\frac{\omega_*}{3}j \!+\! \frac{\omega_*^2}{18}\Delta^2 \!+\! \log  Z^{K, s}_{r_1,r_2}(\omega_\star +i\tau)\right].
\end{align}
The contribution of $(K,r_1,r_2,+)$ solution at $\tau$ equals to the complex conjugate of the contribution of $(K,-r_1,-r_2,-)$ solution at $-\tau$. Thus, we obtain
\begin{align}\label{Y-reflection}
|Y_{j,\Delta} (\tau)|^2 = |Y_{j,\Delta} (-\tau)|^2\ .
\end{align}
In fact, this is an exact symmetry of $Y$ \eqref{SFF}.
In addition, when $\frac{\omega_2}{\omega_1}=\frac{p}{q}$ with coprime integers $p,q$, we find
\begin{align}\label{Y-period}
Y_{j,\Delta} (\tau) \sim Y_{j,\Delta} (\tau+ (p+q) \pi)\ ,
\end{align}
in the large $N$ limit. Hence, $Y$ is periodic only when $\frac{\omega_2}{\omega_1}$ is a rational number.
Combining the above two symmetries \eqref{Y-reflection} and \eqref{Y-period}, we get $|Y_{j,\Delta} (\tau)|^2 = |Y_{j,\Delta} ((p+q)\pi-\tau)|^2$ for $\frac{\omega_2}{\omega_1}=\frac{p}{q}$.

\begin{figure}[htbp]
	\begin{center}
		\includegraphics[width=.48\textwidth]{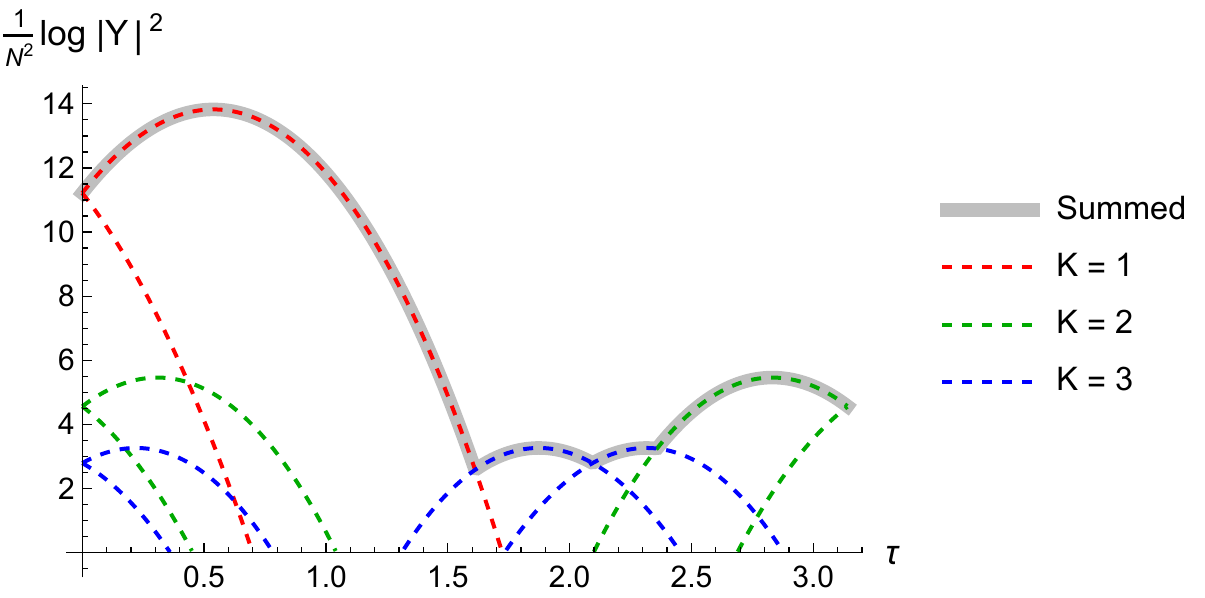}
	\end{center}
	\caption{The plot of the SFF for $\tau \in (0, \pi)$ with $\omega_1=\omega_2,\, N=1000, \, j=10N^2, \, \Delta=10N$. We plot contributions from the $K$-cut saddles and find that each of them dominates in a certain region of $\tau$.}
\label{fig: SFF-1}
\end{figure}

Let us first discuss the results for $\omega_1 = \omega_2$. A typical behavior of the spectral form factor (scaled by $N^2$) for $\omega_1=\omega_2$ and $\Delta/N$ not so large ($\sim 10$) is given in Fig. \ref{fig: SFF-1}.
The highest peaks come from the $K=1$ solutions, while the second and third highest peaks come from $K=2$ and $K=3$ solutions, respectively. The $K \geq 4$ solutions are negligible here. 
For each $K$ and $r=r_1=r_2$, we have two dominant saddles labeled by $s=\pm$, that give rise to two peaks in a symmetric fashion. The center of two peaks coming from the $(K, r)$ saddle is located at $\tau = \frac{2\pi r}{K}$ with $r \in \mathbb{Z}$.

The $K=1$ saddle in the matrix integral corresponds to the BPS black hole of \cite{Gutowski:2004ez} in AdS$_5$. We see that its contribution decays very rapidly in the early time as expected, called the `slope' \cite{Cotler:2016fpe}. Once the $K=1$ contribution decays, the $K>1$ saddles take over, which correspond to the $\mathbb{Z}_K$-orbifolded Euclidean black hole solutions. These new saddles dominantly contribute to the spectral form factor so that it does not decay at late times. 
A similar phenomenon was found in the free large-$N$ Yang-Mills theory \cite{Chen:2022hbi}, where the sub-dominant saddles (at early times) preserving $\mathbb{Z}_K$ subgroup of the center symmetry become dominant at late time. We are tempted to interpret this as an analog of `ramp,' but the linear growth as in the case of Gaussian unitary ensemble (GUE) is somewhat obscure. Moreover, since the `time' of the index is $2\pi$ periodic, it seems that there is no direct analog of the `late-time' behavior for the index version of the SFF. 

\begin{figure}[htbp]
	\begin{center}
		\includegraphics[width=.48\textwidth]{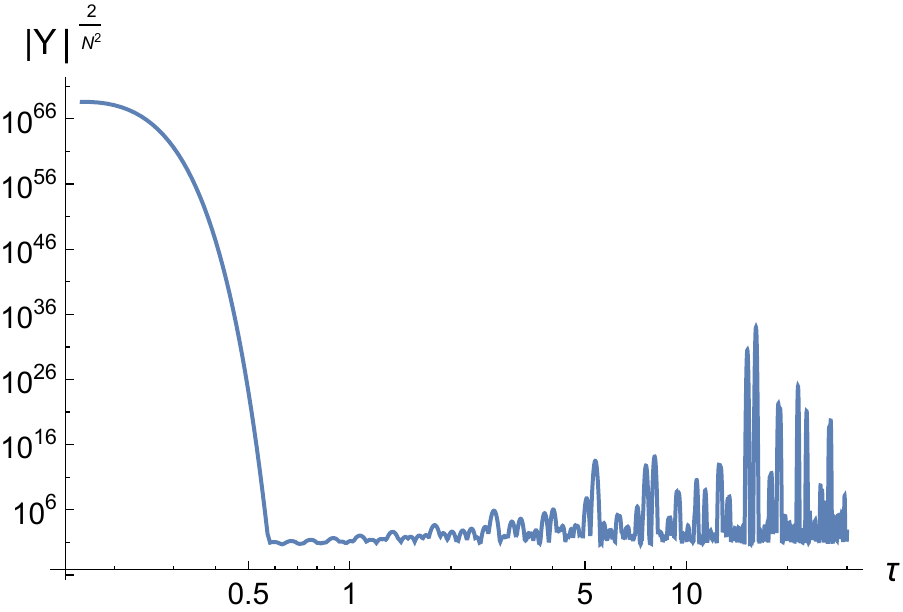}
	\end{center}
	\caption{The log-log plot of the SFF for $\tau \in (0.16,30)$ with $\omega_2 = \sqrt{2}\omega_1, \, N=1000, \, j=100N^2, \, \Delta=100N$.}
\label{fig: SFF-2}
\end{figure}

To this end, we consider the case with $\omega_1 \neq \omega_2$. Especially, if the ratio $c=\omega_1/\omega_2$ is an irrational number, the periodicity of the index is broken. (We may still see an approximate periodicity for $c \simeq p/q$.) Let us choose $\omega_2 = \sqrt{2}\omega_1$ and $\Delta/N$ to be large ($\sim 100$). In this region, a typical behavior of the SFF is given by Fig. \ref{fig: SFF-2}.
We observe that the orbifolded Euclidean black hole solutions become dominant contributions to the SFF in the late time. 
We find that varying $\frac{j}{N^2}$ at fixed $\frac{\Delta}{N}$ does not change much the qualitative behavior of $Y$ and the time scale at which the SFF falls to a minimum (called the dip time in the literature) is of order $\tau_d \sim \frac{N}{\Delta}$ at fixed $\frac{j}{N^2}$. 

Although we do not observe an apparent `plateau' here, these orbifolded black hole solutions are responsible for the increase of the spectral form factor at late times (an analog of the `ramp') for the $\mathcal{N}=4$ SYM theory. We find our analysis to be similar to the case of AdS$_3$/CFT$_2$, where the subdominant saddles at early times become significant at late times \cite{Dyer:2016pou}. 
This is in sharp contrast with the case in the JT gravity or the ensemble-averaged theory, where the Euclidean wormholes are responsible for this behavior \cite{Saad:2018bqo, Saad:2019lba}.

\section{Discussion}

In this paper, we have studied a supersymmetric index analog of the microcanonical spectral form factor and analyzed its property in the large $N$ limit of $\CN=4$ SYM theory. We found that the most dominant saddle describing black hole decays rapidly at early `time.' The subdominant $K$-cut saddles at early times become dominant at late times, so that the SFF rises. 
These saddle points have geometric descriptions in supergravity in the conventional AdS/CFT sense, at least when two chemical potentials are set to be equal $\omega_1 = \omega_2$ \cite{Aharony:2021zkr}. Its free energy is simply given by the on-shell Euclidean action given schematically as $\-S_{\textrm{BH}}/K$. The SFF does not vanish thanks to these orbifolded Euclidean black holes instead of spacetime wormholes. These give rise to an analog of `ramp,' giving a geometric account for the recovery of information.

There are a number of interesting questions that need to be answered. First, it is not yet clear whether these $K$-cut saddles are enough to account for the information loss. Qualitatively, we see that the microcanonical SFF grows, but it is not yet clear whether this is enough. Moreover, there can be other gravity solutions that affect the late-time behavior of SFF. One of the saddle points we have not considered in this paper is the $K$-cut solution with a non-trivial filling fraction \cite{Choi:2021rxi}. 

Also, we do not know if the index version of the SFF can probe chaos. As far as we have observed, it does not seem to exhibit the expected behavior of the random matrix theory. Nevertheless, given our partial success in finding erratic behavior at late times, it is interesting to ask if there is a sign of chaos even in the supersymmetry protected sector. 

It is interesting to notice that the Euclidean black hole solutions associated with the $K$-cut saddles are orbifolded in the entire 10-dimensional spacetime, including the compact $S^5$ direction, not just the AdS$_5$. It may be a hint that the recovery of information can be sensitive to the UV completion of gravity.

Lastly, we have focused on the four-dimensional $\mathcal{N}=4$ SYM theory in this paper. It would be interesting to generalize our discussion to less supersymmetric cases. For the case of 4d $\CN=1$ theory, the superconformal $R$-charges are not necessarily rational numbers. Therefore the index is not necessarily periodic in imaginary chemical potential, which may allow us to probe deeper late time behavior of the SFF. For $\CN=1$ theories, a Cardy-like limit has been studied \cite{Kim:2019yrz, Cabo-Bizet:2019osg}, and a set of sub-dominant saddles (in the sense of free energy) that are similar to our $K$-cut saddles are found \cite{Cabo-Bizet:2019osg}. It is natural to expect there exist saddles for the large $N$ limit of holographic theories that reduce to these in the Cardy limit. It would be also interesting to generalize our discussion in other dimensions (see, for example \cite{Choi:2019zpz,Choi:2019dfu, Nian:2019pxj}, \cite{Choi:2019miv, Crichigno:2020ouj} and \cite{Choi:2018hmj,Nahmgoong:2019hko} for the study of Cardy-like limit in 3d, 5d and 6d SCFTs) and see if similar orbifolding of Euclidean black holes can account for the late time behavior of SFF.

\vspace{6pt}

\begin{acknowledgments}
We thank Monica Kang, Ki-Hong Lee, Kimyeong Lee, Sungjay Lee, and Dario Rosa for helpful discussions. 
SC thanks KAIST for the kind hospitality during his visit, where part of this work was done.
This work is supported by a KIAS Individual Grant (PG081601) at Korea Institute for Advanced Study (SC), the National Research Foundation of Korea (NRF) Grants 2021R1A2C2012350 (SK) and NRF-2020R1C1C1007591 (JS). 
The work of JS is also supported by the POSCO Science Fellowship of POSCO TJ Park Foundation and a Start-up Research Grant for new faculty provided by KAIST.  
\end{acknowledgments}

\bibliography{refs}

\end{document}